\documentclass[twocolumn,showpacs,preprintnumbers,amsmath,amssymb]{revtex4}

\usepackage{graphicx}
\usepackage{dcolumn}
\usepackage{bm}

\begin{document}

\preprint{APS/123-QED}

\title{Pumping in quantum dots and non-Abelian matrix  Berry phases}
\author{N.Y. Hwang}
\author{S.C. Kim}
\author{P.S. Park}
\author{S.-R. Eric Yang$\footnote{ corresponding author, eyang@venus.korea.ac.kr}$}
\affiliation{Physics Department, Korea  University, Seoul Korea }

\date{\today}

\begin{abstract}
We have investigated    pumping in quantum dots from the perspective of
non-Abelian (matrix) Berry phases by
solving the time dependent Schr{\"o}dinger equation exactly for adiabatic changes.
Our results demonstrate that a pumped charge is  related to the presence of  a  finite matrix Berry phase.
When consecutive
adiabatic cycles are performed
the pumped charge of each cycle is different from the previous ones.
\end{abstract}

\pacs{71.55.Eq, 71.70.Ej, 03.67.Lx, 03.67.Pp}
\maketitle

\section{Introduction}

\begin{figure}[!hbt]
\begin{center}
\includegraphics[width = 0.4 \textwidth]{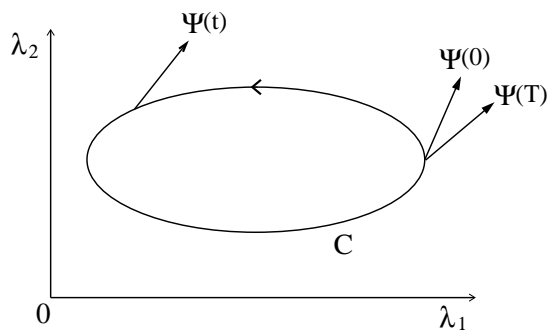}
\caption{Adiabatic  path in the parameter space is shown.
An electron is in the state $|\Psi(0)\rangle$ initially.
After the pumping action it is in the final state $|\Psi(T)\rangle$, where $T$ is the period of the cycle.
The initial and final states
$|\Psi(0)\rangle$ and $|\Psi(T)\rangle$ are related to each other through a matrix Berry phase.
During the adiabatic cycle the electron state does not leave the doubly degenerate Hilbert subspace.
}
\label{fig:ad}
\end{center}
\end{figure}

A semiconductor quantum dot electron  pump\cite{Th, Alt} is a device
that  can make a dc current flow in unbiased systems. In such a
device a charge can be pumped through the dot when the shape  of the
double barrier  potential undergoes an adiabatic cyclic change.
Quantum pumping is a fascinating subject and considerable
experimental\cite{Sw,Ko,Pot,Wat} and theoretical studies have been
carried out\cite{Br,Zh,Sh,An,Al,Av,OE,Zho,Ma,Lev,Tor,Ag2,Se,Be}.
Recently effects of noise and decoherence\cite{Levit,Mos1,Mos2,Pol},
interactions\cite{Al2,Br2,Ao,Ci,Sp,Shar,Ag}, and
spin\cite{Wat,Mu,Go,Bl} have been investigated. The pumped charge
through a single  energy level during a cycle  at zero temperature
can be calculated\cite{Br} using emissivity\cite{Bu}
\begin{eqnarray}
Q=\frac{e}{\pi}\int_A \mathrm d \lambda_1 \mathrm d \lambda_2
\mathrm{Im} \left( \frac{\partial s_{11}^*}{\partial \lambda_1} \frac{\partial s_{11}}{\partial \lambda_2}
+\frac{\partial s_{12}^*}{\partial \lambda_1} \frac{\partial s_{12}}{\partial \lambda_2} \right),
\label{eq:scatt}
\end{eqnarray}
where $s_{\alpha \beta}$ is the scattering matrix element, and $\alpha$ and $\beta$ denote contacts $1$ and $2$.
During an adiabatic cycle the parameters $\lambda_1$ and $\lambda_2$, which control the shape of the dot,
follow a closed path $C$ in the parameter space and the area defined by it is denoted as  $A$,  see Fig.\ref{fig:ad}.

The integration over the area $A$
suggests  that quantum pumping depends on the geometric properties
of the path and is independent on detailed time evolution.
This indicates  that
quantum pumping in nano-semiconductors may be  a manifestation of  a matrix (non-Abelian) Berry phase\cite{Wil,Sha}.
Similar issues have been investigated in quite different physical systems, such as
superconducting circuits \cite{Bro}.
Recent investigations have shown that several semiconductor dots  possess non-Abelian matrix  Berry phases.
Semiconductor
excitons\cite{Sol} and p-type electrons\cite{Sere2,Bern} can  exhibit matrix Berry phases.
It has also been shown that  n-type semiconductor dots  with spin-orbit coupling
possess a matrix (non-Abelian) Berry phase\cite{yang1,yang2,yang3}.
This is due to the presence of energy level degeneracy originating from  time-reversal symmetry.

\begin{figure}[!hbt]
\begin{center}
\includegraphics[width = 0.35 \textwidth]{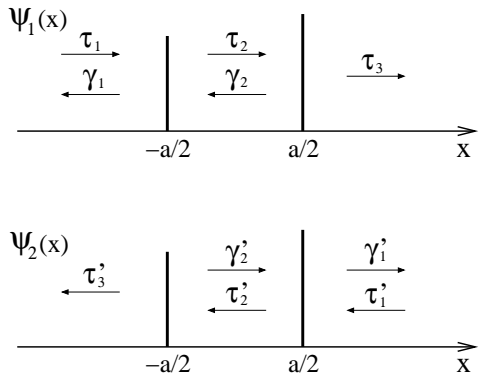}
\caption{In our model an electron moves in a one-dimensional asymmetric double barrier system without spin-orbit coupling.
Two degenerate scattering states $\psi_1(x)$ and $\psi_2(x)$ are shown.  $\tau_i$, $\tau^{'}_i$ and $\gamma_i, \gamma^{'}_i$ are wavefunction amplitudes.}
\label{fig:setup1}
\end{center}
\end{figure}

We show in this paper that  adiabatic pumping  in semiconductor   quantum dots without spin-orbit coupling
may be analyzed in terms of non-Abelian
Berry phases.
For  non-Abelian
Berry phases to be finite  it  is essential that  degenerate states are present, and semiconductor quantum pumps
do have  non-trivial degenerate scattering states, see Fig.\ref{fig:setup1}.
Each energy level of  a quantum dot pump  is doubly degenerate since there are two independent
scattering states incident from the left and  right  barriers.
During the pumping action
these degenerate states can get mixed.
This
can be seen as follows: Let $T(t+\delta t,t)$ be the time development operator between $t+\delta t$ and $t$, and
let the change of the confinement potential of the dot  during the pumping action  be $\delta V(t)=V(t+\delta t)-V(t)$, where
$\delta t$ is an infinitely small time interval.
The  degenerate  instantaneous eigenstates of the Hamiltonian are chosen as the degenerate
scattering states $|\psi_i(t)\rangle $, shown in Fig.\ref{fig:setup1}.
Suppose that the initial electron at $t$ is in the scattering state $|\psi_1(t)\rangle$.  Then the overlap matrix element of
the time development operator between
the degenerate   scattering eigenstates of the Hamiltonian is given by
\begin{eqnarray}
\langle \psi_2(t+\delta t,t)  |T(t+\delta t,t) |\psi_1(t)\rangle \approx \nonumber\\
\frac{i\delta t}{\hbar}\langle \psi_2(t) |\delta V(t) |\psi_1(t)\rangle
\label{eq:linear}
\end{eqnarray}
since $T(t+\delta t,t) \approx 1-\frac{i}{\hbar}\delta t H(t)$.
For some pumping actions, given by Eq.(\ref{closed_loop}), this overlap is {\it finite}, and
the state $|\psi_1(t)\rangle $  can {\it develop} into a linear combination
of  $|\psi_1(t+\delta t)\rangle $ and  $|\psi_1(t+\delta t)\rangle $.
This suggests that during the adiabatic cycle a state vector may rotate in the degenerate Hilbert subspace
and the initial state  may  not return to
itself after the adiabatic cycle:  $|\Psi(0)\rangle\neq |\Psi(T)\rangle$, as shown in  Fig.\ref{fig:ad}.
It is unclear under what conditions this happens and how different the initial and final states  are.
Do they differ  by a phase factor or more than a phase factor?
Theory of non-Abelian matrix Berry phases\cite{Wil} is well suited to  address these  issues.

We show that the presence of a finite matrix Berry phase leads
to a non-zero pumped charge.
Our investigation of time evolution of the electron state shows that
the pumped charge depends on the initial state of the adiabatic cycle, i.e., different initial states
give different pumped charges.
Our method is applicable even when the potential modulations are substantial.

\section{Non-Abelian U(2) gauge theory, matrix Berry phase, and pumped charge}

Non-Abelian Berry phases\cite{Wil} are computed from the  {\it exact} solutions of
the time-dependent Schr{\"o}dinger equation for adiabatic changes.
In this approach
one can choose the  orthonormal basis states of
the degenerate energy level as the instantaneous eigenstates of the Hamiltonian:
$H(t)|\Phi_i(t)\rangle=E(t)|\Phi_i(t)\rangle$ for  $i=1,2$.
(Here we consider only adiabatic changes and other states with different energies are ignored).
Even when the electron state starts in one of the instantaneous eigenstates $|\Phi_i(0)\rangle$ it may not stay
in the same eigenstate $|\Phi_i(t)\rangle$ at later times:
the actual state at $t$ may be  a {\it linear combination} of the instantaneous eigenstates
\begin{eqnarray}
|\Psi(t)\rangle =c_1(t)|\Phi_1(t)\rangle +c_2(t)|\Phi_2(t)\rangle.
\label{eq:instan}
\end{eqnarray}
For a cyclic change with the  period  $T$,
the instantaneous basis states $|\Phi_1(T)\rangle$ and  $|\Phi_2(T)\rangle$ return to the initial states $|\Phi_1(0)\rangle$ and
$|\Phi_2(0)\rangle$, but the coefficients
$c_1(T)$ and $c_2(T)$ may not return to the initial values.
In this case,
the initial and final  degenerate states may be  connected via
a {\it finite} matrix Berry phase.
A matrix  Berry phase is a non-perturbative concept and is applicable
even in the presence of strong  potential modulations that are adiabatic.
A matrix Berry  phase can arise when an energy level has a {\it non-trivial degeneracy} which does not split
during an adiabatic change of the
external parameters $\lambda_p$ of the Hamiltonian.
For a doubly degenerate energy level a   matrix Berry phase is  a $2\times2$ matrix
$\Phi_C$ connecting the final amplitudes $(c_1(T), c_2(T))$ to initial  amplitudes $( c_1(0),c_2(0))$:
\begin{eqnarray}
\left(
\begin{array}{c}
c_1(T) \\
c_2(T)
\end{array}
\right)=\Phi_C
\left(
\begin{array}{c}
c_1(0) \\
c_2(0)
\end{array}\right)
\label{eq:matrixBerry}.
\end{eqnarray}
Using the instantaneous degenerate basis vectors the matrix Berry phase can be computed by solving  the
adiabatic version of the time-dependent Schr{\"o}dinger equation
\begin{eqnarray}
i \hbar \dot{c}_i=-\sum_j A_{i j} c_j \qquad i=1,2,
\label{eq:time_Schrod}
\end{eqnarray}
where $A_{ij}= \hbar \sum_p(A_p)_{i,j}\frac{d\lambda_p}{dt}$ and  the
sum over $p$ in $A_{ij}$ is meant to be the sum  over $\lambda_p$.
The non-Abelian gauge potentials are defined as
\begin{eqnarray}
(A_p)_{i,j}=
i \langle\Phi_i|\frac{\partial\Phi_j}{\partial\lambda_p}\rangle .
\label{eq:vec}
\end{eqnarray}

Solving  the time-dependent Schr{\"o}dinger equation,
Eq.(\ref{eq:time_Schrod}),  one can write the matrix Berry phase in
terms of the non-Abelian gauge potentials\cite{Wil}:
\begin{eqnarray}
\Phi_C
&=&e^{ i \sum_p A_p(t_n) d\lambda_p}....e^{ i  \sum_pA_p(t_1) d\lambda_p}\nonumber\\
&=&Pe^{ i \oint_C\sum_p A_p d\lambda_p},
\label{eq:matB}
\end{eqnarray}
where  a path ordering $P$ must be used since $A_p$ at different times are usually noncommuting.
We see from this contour integration  that
\begin{enumerate}
\item
Matrix Berry phase is a geometric effect, i.e. it depends  only  on the path C.
\item
Matrix Berry phase  is independent of the period of the cycle as long as the change is adiabatic.
\end{enumerate}
It is instructive to compare this result to  the Abelian Aharonov-Bohm phase of
an electron in the electromagnetic vector potential $ \vec{A}$:
\begin{eqnarray}
\Phi_C=e^{ i \oint_C \vec{A} \cdot d\vec{\ell}},
\end{eqnarray}
where $C$ is a closed path encircling the magnetic flux.
The only difference between the Abelian and non-Abelian phases is the mathematical nature of the vector potentials:
a number vs. matrix.
Under a unitary transformation
$| \psi'_i \rangle =\sum_j U^*_{i j}| \psi_j \rangle$
the non-Abelian gauge structure emerges
\begin{eqnarray}
A'_k=UA_kU^\dagger+i U \frac{\partial U^\dagger}{\partial \lambda_k}.
\label{eq:vec_pot}
\end{eqnarray}
For doubly degenerate levels the relevant {\it gauge group} is  $U(2)$.

A pumped charge can be computed from the solution of  the
time-dependent Schr{\"o}dinger equation, Eq.(\ref{eq:time_Schrod}).
It  can be calculated from  the  total probabilities   in
the regions
to the left, right, and center of the dot, denoted by
\begin{eqnarray}
P_\mathrm L(t)&=& \int_{-L}^{-a/2}|\Psi(x,t)|^2 \mathrm d x,\nonumber\\
P_\mathrm R(t)&=& \int_{a/2}^{L}|\Psi(x,t)|^2 \mathrm d x,\nonumber\\
P_\mathrm C(t)&=& \int_{-a/2}^{a/2}|\Psi(x,t)|^2 \mathrm d x,
\label{eq:pumpch}
\end{eqnarray}
where $L$ is the length of the system.
The change of these probabilities after the $n$th  adiabatic cycle, $t=nT$, is
\begin{eqnarray}
\Delta P_\mathrm i(n)=P_\mathrm i(n)-P_\mathrm i(n-1).
\end{eqnarray}
for  $i=L,C,R$. The probabilities $\Delta P_L$ and  $\Delta P_R$ are
directly related to the pumped charge: the pumped charge after the
first cycle is
\begin{eqnarray}
Q=\Delta P_L(1)=-\Delta P_R(1),
\label{eq:pumped_charge}
\end{eqnarray}
provided that the change of probability in the dot, $\Delta P_C(1)$, is negligible.
It is straightforward  to relate the matrix Berry phase
to the pumped charge using Eqs. (\ref{eq:matrixBerry}) and (\ref{eq:pumped_charge}).
The dependence on the non-Abelian Berry phases enters through Eq.(\ref{eq:matrixBerry}),
and the pumped charge  depends on the initial
conditions $(c_1(0), c_2(0))$ and
basis states $\Phi_j(x)$:
\begin{eqnarray}
Q&=&\sum_{ij}O_{ij}[c_i(T)^*c_j(T)-c_i(0)^*c_j(0)]\nonumber\\
&=&\sum_{ijkl}O_{ij}[(\Phi_C)^*_{ik}(\Phi_C)_{jl}-\delta_{ik}\delta_{jl}]c_k(0)^*c_l(0),\nonumber\\
\end{eqnarray}
where
\begin{eqnarray}
O_{ij}&=&\int_{-L}^{-a/2}\Phi_i(x)^*\Phi_j(x)dx.
\end{eqnarray}

\section{Model calculation of pumped charge}

\  \\
\  \\
\begin{figure}[!hbt]
\begin{center}
\includegraphics[width = 0.4 \textwidth]{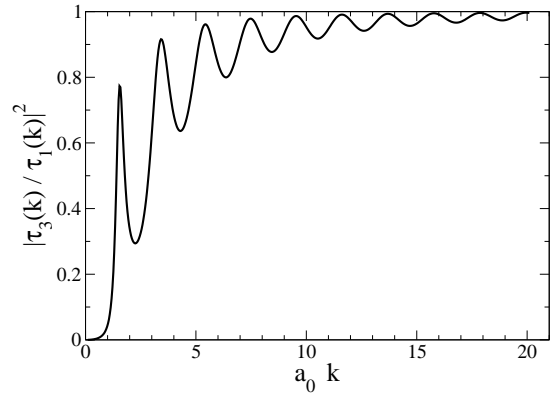}
\caption{Transmission coefficient of a  double barrier system with the parameters
$V_l/(E_0a_0)=4.25$ and $V_r/(E_0a_0)=2.0$.}
\label{fig:transmission}
\end{center}
\end{figure}

The simplest possible model to study quantum pumping is a one-dimensional double barrier system\cite{Bl1,Wei}.
If the barrier potentials are chosen to be delta functions
it is possible to calculate analytically the eigenstates and the non-Abelian vector
potentials.
Let us consider two delta-function barriers with the distance $a$:
\begin{eqnarray}
V(x)=V_{l} \delta(x+\frac{a}{2})+V_{r} \delta(x-\frac{a}{2}),
\label{eq:barriers}
\end{eqnarray}
where $V_{l}$ and $V_{r}$ are heights of the left and right
barriers, respectively. Each energy level is {\it doubly degenerate}
with  eigenstates  $\psi_1(x)$ and  $\psi_2(x)$: in $\psi_1(x)$
($\psi_2(x)$) the electron is incident from the left (right), see
Fig.1. Due to time reversal invariance the scattering amplitudes
satisfy $\tau_3/\tau_1=\tau_3'/\tau_1'$. In the following we
consider the pumped charge of an electron in the energy level
$E=\frac{\hbar^2k^2}{2m}$, where $k$ is a wavevector and $m$ is the
electron effective mass. We use the energy unit
$E_0=\frac{\hbar^2}{2m a_0^2}$ and the length unit  $a_0=2a/3$.

In the numerical calculation of the pumped charge it is convenient
to use box normalization with length $L$. Let us  comment on how the
limit $L\rightarrow \infty$ is taken. It is natural to choose  the
incoming scattering states from the left and  right as the
instantaneous  basis states: $\psi_1(x)$ and $\psi_2(x)$ (any other
orthonormal set can also be used to calculate the non-Abelian vector
potentials). However, for a finite value of $L$ these two states are
not orthogonal. We  choose one of the instantaneous basis state as
$\psi_1(x)$ and construct the other basis state via Gram-Schmidt
orthogonalization using  $\psi_2(x)$:
\begin{eqnarray}
&&\Phi_1(x)=\psi_1(x),\nonumber\\
&&\Phi_2(x)=\alpha\psi_1(x)+\beta\psi_2(x),
\label{eq:basis}
\end{eqnarray}
As $L \rightarrow  \infty $ the overlap $\langle\psi_1|\psi_2\rangle \rightarrow 0$ and  $\Phi_2(x)\rightarrow \psi_2(x)$.
Below the pumped charge is calculated for sufficiently large values $L$ where the size dependence is negligible.

\subsection{Parity symmetry: absence of matrix Berry phase  and pumped charge}

When the double barrier potential retains parity symmetry during the adiabatic cycle, i.e. $V_l(t)=V_r(t)$,
the pumped charge is zero.
(The adiabatic parameters are $(V_l(t),a(t))$).
This follows from the expression for the pumped charge given in Eq.(\ref{eq:scatt}):
we have verified that the imaginary part of
$\frac{\partial s_{11}^*}{\partial V_l} \frac{\partial s_{11}}{\partial a}
+\frac{\partial s_{12}^*}{\partial V_l} \frac{\partial s_{12}}{\partial a}$ is zero.
We show that this is fully {\it consistent} with the absence  of a matrix Berry phase.
For this purpose it is convenient to choose the following orthonormal instantaneous basis states:
\begin{eqnarray}
&&\Phi_1(x)=\frac{1}{\sqrt{N_1}}(\psi_1(x)+\psi_2(x)),\nonumber\\
&&\Phi_2(x)=\frac{1}{\sqrt{N_2}}(\psi_1(x)-\psi_2(x)),
\label{eq:basis2}
\end{eqnarray}
where the normalization factors are
$N_1=2+2\langle\psi_1|\psi_2\rangle$ and
$N_2=2-2\langle\psi_1|\psi_2\rangle$.
We will use this orthonormal set to calculate the non-Abelian vector potentials\cite{com2}.
Since  two barrier heights are
equal, i.e. $V_l=V_r$, it can be easily shown that
$\psi_1(x)=\psi_2(-x)$.
From this it follows that $\Phi_1(x)$  and $\Phi_2(x)$ are, respectively, {\it even} and {\it odd} functions in $x$.
Taking the above states as basis vectors,
and using the results
{\setlength\arraycolsep{2pt}
$\langle\psi_1|\frac{\partial}{\partial
\lambda_p}|\psi_1\rangle
=\langle\psi_2|\frac{\partial}{\partial \lambda_p}|\psi_2\rangle$
and  $\langle\psi_1|\frac{\partial}{\partial\lambda
p}|\psi_2\rangle=\langle\psi_2|\frac{\partial}{\partial\lambda p}|\psi_1\rangle$
with the orthogonality $\langle\Phi_1|\Phi_2\rangle=0$
we find that
the following  off-diagonal elements of vector potentials are zero:
{\setlength\arraycolsep{2pt}
\begin{eqnarray}
(A_p)_{12}&=&i \langle\Phi_1|\frac{\partial\Phi_2}{\partial\lambda_p}\rangle\nonumber\\
&=& \frac{i}{\sqrt{N_1^{*}N_2}}\big(\langle\psi_1|\frac{\partial}{\partial\lambda_p}|
\psi_1\rangle+\langle\psi_1|\frac{\partial}{\partial\lambda_p}|\psi_2\rangle\nonumber\\
&&-\langle\psi_2|\frac{\partial}{\partial\lambda_p}|\psi_1\rangle-\langle\psi_2|\frac{\partial}{\partial\lambda_p}|\psi_2\rangle\big)=0.
\end{eqnarray}}
Similarly $(A_p)_{21}=0$ can be proved.
Since the off-diagonal elements of the vector potentials are zero the matrix Berry phase is zero.
This effect of parity symmetry is already known from the investigation of matrix Berry phases of isolated II-VI and III-V quantum dots
in the presence of spin-orbit coupling:  when the in-plane confinement
potential of the dot has parity symmetry the matrix Berry phase vanishes\cite{yang1,yang2}.

\subsection{Numerical calculation of pumped charge}

\begin{figure}[!hbt]
\begin{center}
\includegraphics[width=0.5 \textwidth]{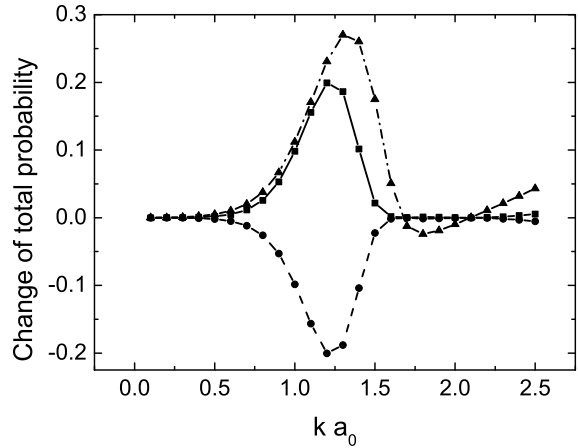}
\caption{ Numerical results for the change in the  total probabilities in different space regions after the first adiabatic cycle
as a function of wavevector:
$\Delta P_\mathrm L(1)$ (squares) and  $\Delta P_\mathrm R(1)$ (circles).
The pumped charge $Q/e$ calculated analytically using Eq.(\ref{eq:scatt})(triangles).  We have used $L/a_0=500$ and
$\omega/E_0=0.1$.
}
\label{fig:reson}
\end{center}
\end{figure}

When the  parity symmetry is broken during the adiabatic cycle, i.e. $V_l(t)\neq V_r(t)$,
a pumped charge is expected.
We choose the adiabatic parameters as $V_l$ and $V_r$ and select elliptic paths
\begin{eqnarray}
&&(V_l(t),V_r(t))=\nonumber\\
&&(V_{l,c}+\Delta V_l \cos(\omega t), V_{r,c}+\Delta V_r
\sin(\omega t)).
\label{closed_loop}
\end{eqnarray}
At $t=0$ we start with the scattering  state $|\psi_1\rangle$. We
investigate pumped charges when the potential modulations are {\it
substantial} : $\Delta V_l/V_{l,c}=0.7$, $\Delta V_r/V_{r,c}=0.7$,
where  $V_{l,c}/(a_0E_0)=2.5$ and   $V_{r,c}/(a_0E_0)=2.0$. At the
start of the adiabatic cycle the quantum dot is characterized by the
scattering amplitudes shown in Fig.\ref{fig:transmission}. The time
dependent Schr{\"o}dinger  equation, Eq.(\ref{eq:time_Schrod}) is
solved numerically, and we determine the amplitudes $c_{1,2}(T)$,
from which the matrix Berry phase can be computed using
Eq.(\ref{eq:matrixBerry}).  The pumped charge after the first cycle,
$\Delta P_\mathrm L(1)$, is calculated as a function of wavevector
using Eqs.(\ref{eq:pumpch}). The results are shown in
Fig.\ref{fig:reson}. The pumped charge is maximum near $ka_0=1.25$,
which is close to the position of the first resonance in the
transmission coefficient, $ka_0=1.56$\cite{Wei,Bl1}. Note that  the
vanished probability on the right side of the dot appears on the
left side: $\Delta P_\mathrm L(1)= -\Delta P_\mathrm R(1)$. This
implies that the pumped charge does not accumulate in the dot, which
is consistent with  Eq.(\ref{eq:scatt}), see Ref.\cite{Br}. We have
also compared quantitatively our results for the pumped charges with
those obtained from Eq.(\ref{eq:scatt}): they  are in a good
qualitative agreement.

\begin{figure}[!hbt]
\begin{center}
\includegraphics[width=0.5 \textwidth]{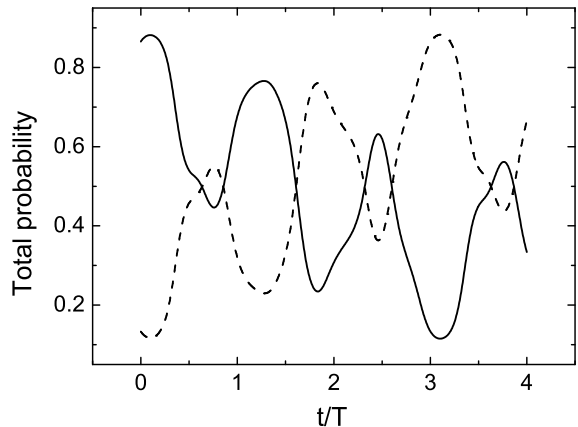}
\caption{Time dependence of the  total probabilities
in different space regions, $P_\mathrm L$ (solid) and  $P_\mathrm R$ (dashed),
during  consecutive adiabatic transformations.
}
\label{fig:num_elec}
\end{center}
\end{figure}

In our approach we can follow microscopically how various quantities
actually depend on time.
Fig.\ref{fig:num_elec} displays  time dependence of the
total probabilities in different space regions during  consecutive
adiabatic transformations. Results are shown for  $k a_0=1.30$. We
note that  $P_L(n)$ for $n=0,1,...$ are all different, implying that
pumped charge $\Delta P_L(n)$ depends on the number of consecutive
adiabatic cycles.   This is because initial  state of the $n$th
cycle $\Psi((n-1)T)$ is  different from that of the $(n-1)$th cycle
$\Psi((n-2)T)$.
In Eq.(\ref{eq:scatt}) this dependence on initial states is not present.  We believe it  is because  the initial states are
chosen to be some particular scattering states, see, for example, Eq.(\ref{eq:pumped_charge}) of Ref.\cite{OE}.
Note that the probability fluxes through
the left and right barriers are equal to each other at all times:
$\frac{dP_L}{dt}= -\frac{dP_R}{dt}$. During the  period of  a cycle
the instantaneous pumped current displays a {\it significant}
variation in time.

\section{ Conclusions}

Solving the time dependent Schr{\"o}dinger equation exactly for
adiabatic changes we find  that the initial and final states are
connected via a {\it finite} matrix Berry phase.  This  difference
between initial and final states of an adiabatic cycle implies that
a pumped charge is present. The main physics behind this is that
even when the electron state starts in one of the instantaneous
eigenstates it may  not stay in the same eigenstate  at later times:
often the actual state at time $t$ turns out to be  a {\it linear
combination} of the instantaneous eigenstates. One can  prepare the
initial state to be the incident scattering state from the left.
When  consecutive adiabatic cycles are performed starting from this
initial state the pumped charge at the $n$th cycle is not
necessarily the same as that of the $(n-1)$th cycle. Our
investigation of the time dependence thus shows that the pumped
charge depends on the nature of the initial state of a cycle.
Moreover, during the   period of
a cycle  the probability flux through the dot  displays a
significant variation in time. It would be interesting to test these
results experimentally.

\section*{Acknowledgments}
S.R.E.Y thanks M.Governale for useful discussions. He is also
grateful to R.Fazio for reading this manuscript and bringing
Ref.[38] to his attention. This work was  supported by grant No.
R01-2005-000-10352-0 from the Basic Research Program of the Korea
Science and Engineering Foundation and by Quantum Functional
Semiconductor Research Center (QSRC) at Dongguk University of the
Korea Science and Engineering Foundation. This work was supported by
The Second Brain Korea 21 Project.


\begin{thebibliography}{00}
\bibitem{Th} D.J. Thouless, Phys. Rev. B {\bf 27}, 6083 (1983).
\bibitem{Alt} B. L. Altschuler and L. I. Glazman, Science {\bf 283}, 1864 (1999).
\bibitem{Sw} M. Switkes, C. M. Marcus, K. Campman and A. C. Gossard, Science {\bf 283}, 1905 (1999).
\bibitem{Ko} L.P.Kouwenhoven, A. T. Johnson, N. C. vanderVaart, C. J. P. M. Harmans, C. T. Foxon, Phys. Rev. Lett. {\bf 67}, 1626 (1991).
\bibitem{Pot} H. Pothier, P. Lafarge, C. Urbina, D. Esteve and M. H. Devoret, Europhys. Lett {\bf 17}, 249 (1992).
\bibitem{Wat} S. K. Watson, R.M. Potok, C.M. Marcus, and V. Umansky, Phys. Rev. Lett.{\bf 91}, 258301 (2003).
\bibitem{Br} P. W. Brouwer, Phys. Rev. B {\bf 58}, R10135 (1998).
\bibitem{Zh} F. Zhou, B. Spivak and B. Altshuler, Phys. Rev. Lett. {\bf 82}, 608 (1999).
\bibitem{Sh} T.A. Shutenko, I.L. Aleiner, and B.L. Altshuler, Phys. Rev. B {\bf 61}, 366 (2000).
\bibitem{An} A. Andreev and A. Kamenev, Phys. Rev. Lett. {\bf 85}, 1294 (2000).
\bibitem{Al} I.L. Aleiner, B.L. Altshuler, and A. Kamenev, Phys. Rev. B {\bf 62}, 10373 (2000).
\bibitem{OE} O. Entin-Wohlman, A. Aharony and Y. Levinson, Phys. Rev. B {\bf 65}, 195411 (2002).
\bibitem{Av} J. E. Avron, A. Elgart, G. M. Graf and L. Sadun,  Phys. Rev. B {\bf 62}, R10618 (2000).
\bibitem{Zho}H.-Q. Zhou, S. Y. Cho, and R. H. McKenzie, Phys. Rev. Lett.{\bf 91}, 186803 (2003).
\bibitem{Ma} Y. Makhlin and A. D. Mirlin,  Phys. Rev. Lett.{\bf 87}, 276803 (2001).
\bibitem{Lev} Y. Levinson, O. Entin-Wohlman and P. Wolfle, Physica A {\bf 302}, 335 (2001).
\bibitem{Tor} L.E.F. Foa Torres, Phys. Rev. B {\bf 72}, 245339 (2005).
\bibitem{Ag2} A. Agarwal and D. Sen, J. Phys. Condens. Matter {\bf 19}, 046205 (2007).
\bibitem{Se} I. Sela and D. Cohen, J. Phys. A: Math. Gen. {\bf 39}, 3575 (2006).
\bibitem{Be} M. V. Berry and J. M. Robbins, Proc. R. Soc. Lond. A {\bf 442}, 659 (1993).
\bibitem{Levit} L. S. Levitov, arXiv:cond-mat/0103617.
\bibitem{Mos1} M. Moskalets and M. B{\"u}ttiker, Phys. Rev. B {\bf 66}, 035306 (2002).
\bibitem{Mos2} M. Moskalets and M. B{\"u}ttiker, Phys. Rev. B {\bf 64}, 201305(R) (2001).
\bibitem{Pol} M.L. Polianski, M. G. Vavilov and P.W. Brouwer, Phys. Rev. B {\bf 65}, 245314 (2002).
\bibitem{Al2} I. L. Aleiner and A. V. Andreev, Phys. Rev. Lett. {\bf 81}, 1286 (1998)
\bibitem{Br2} P. W. Brouwer, A. Lamacraft, and K. Flensberg, Phys. Rev. B {\bf 72}, 075316 (2005).
\bibitem{Ao} T. Aono, Phys. Rev. Lett. {\bf 93}, 116601 (2004).
\bibitem{Ci} R. Citro, N. Andrei, and Q. Niu, Phys. Rev. B {\bf 68}, 165312 (2003)
\bibitem{Sp} J. Splettstoesser, M. Governale, J. K{\"o}nig, and R. Fazio, Phys. Rev. B {\bf 74}, 085305 (2006)
\bibitem{Shar} P. Sharma and C. Chamon, Phys. Rev. Lett. {\bf 87}, 96401 (2001)
\bibitem{Ag} A. Agarwal and D. Sen, Phys. Rev.B 76,035308(2007).
\bibitem{Mu} E. R. Mucciolo, C. Chamon, and C. M. Marcus, Phys. Rev. Lett. {\bf 89}, 146802 (2002)
\bibitem{Go} M. Governale, F. Taddei, and R. Fazio, Phys. Rev. B {\bf 68}, 155324 (2003)
\bibitem{Bl} M. Blaauboer, Phys. Rev. B {\bf 68}, 205316 (2003).
\bibitem{Bu} M. B{\"u}ttiker, H. Thomas, and  A. Pr\^{e}tre, Z. Phys. B,  {\bf 94}, 133 (1994).
\bibitem{Wil}F. Wilczek and A. Zee, Phys. Rev. Lett. {\bf 52}, 2111 (1984);
\bibitem{Sha}{\it Geometric Phases in Physics}, edited by A. Shapere and F. Wilczek
(World Scientific, Singapore, 1989).
\bibitem{Bro} V. Brosco, R. Fazio, F.W.J. Hekking, and A. Joye, arXiv:cond-mat/0702333.
\bibitem{Sol} P.Solinas, P. Zanardi, N. Zanghi, F. Rossi, Phys. Rev. A {\bf 67}, 062315 (2003).
\bibitem{Sere2}Yu. A. Serebrennikov, Phys. Rev. B {\bf 70}, 064422 (2004).
\bibitem{Bern}B. A. Bernevig and S.-C. Zhang, Phys. Rev. B {\bf 71}, 035303 (2005).
\bibitem{yang1}S.-R. Eric Yang and N.Y. Hwang, Phys. Rev. B {\bf 73}, 125330 (2006).
\bibitem{yang2}S.-R. Eric Yang, Phys. Rev. B {\bf 74}, 075315 (2006).
\bibitem{yang3}S.-R. Eric Yang,  Phys. Rev. B  {\bf 75}, 245328 (2007).
\bibitem{Bl1} M. Blaauboer and E. J. Heller, Phys. Rev. B {\bf 64}, 241301(R) (2001).
\bibitem{Wei} Y. Wei, J. Wang, and H. Guo, Phys. Rev. B {\bf 62}, 9947 (2000).
\bibitem{com2}
As long as a degenerate pair of basis states can be related to
another pair by a unitary transformation the non-Abelian gauge
structure, Eq.(\ref{eq:vec_pot}), is satisfied, see Ref.\cite{Wil}.
This is true independent of any model.



\end{thebibliography}
\end{document}